\documentstyle[aps,prb]{revtex}
\begin{document}

\input epsf.sty
\twocolumn[\hsize\textwidth\columnwidth\hsize\csname %
@twocolumnfalse\endcsname

\draft

\widetext

\title{Electronic phase separation in lightly-doped
La$_{2-x}$Sr$_x$CuO$_4$}
\author{M. Matsuda}
\address{
Advanced Science Research Center,
Japan Atomic Energy Research Institute, Tokai, Ibaraki 319-1195, Japan}
\author{M. Fujita and K. Yamada}
\address{
Institute for Chemical Research, Kyoto University, Gokasho, Uji
610-0011, Japan}
\author{R. J. Birgeneau}
\address{
Department of Physics, University of Toronto, Toronto, Ontario M5S 1A1, Canada}
\author{Y. Endoh}
\address{
Institute for Materials Research, Tohoku University, Katahira,
Sendai 980-8577, Japan}
\author{G. Shirane}
\address{
Department of Physics, Brookhaven National Laboratory, Upton, New
York 11973}

\date{November 7, 2001}
\maketitle
\begin{abstract}
The hole concentration dependence of the magnetic correlations is
studied in very lightly-doped La$_{2-x}$Sr$_x$CuO$_4$ ($x<$0.02), which
shows coexistence of a three-dimensional antiferromagnetic (AF)
long-range ordered phase and a spin-glass phase at low temperatures.
It is found that the spin-glass phase in the coexistence region also
shows a diagonal spin modulation as in the pure spin-glass phase in
La$_{2-x}$Sr$_x$CuO$_4$ (0.02$\le x\le$0.055).
Below $x\sim$0.02 the diagonal stripe structure is the same
as that in $x\sim$0.02 with a volume fraction almost proportional to hole
concentration, suggesting that electronic phase separation
of the doped holes occurs so that some regions with hole
concentration $c\rm_h\sim$0.02 and the rest with $c\rm_h\sim$0 are formed.
This represents the most direct observation to-date of electronic
phase separation in lightly-doped antiferromagnets. Such phase
separation has been predicted by a number of theories.
\end{abstract}

\phantom{.}
]
\narrowtext

\section{Introduction}
Extensive neutron elastic scattering studies on lightly-doped
La$_{2-x}$Sr$_x$CuO$_4$ have revealed that a diagonal spin modulation,
which is a one-dimensional modulation rotated away by 45$^\circ$ from
that in the superconducting phase, occurs universally across the
spin-glass phase in La$_{2-x}$Sr$_x$CuO$_4$
(0.02$\le x\le$0.055).~\cite{wakimoto,wakimoto2,matsudanew,fujita}
Such diagonal stripes have been
predicted theoretically \cite{machida0,kato,rice,schulz,zaanen}
and have also been observed in insulating
La$_{2-x}$Sr$_x$NiO$_4$. \cite{tranquada3,yoshizawa}
These results lead to the important
conclusion that the static magnetic spin modulation changes from being
diagonal to parallel at $x=0.055\pm0.005$, coincident with the
insulator-to-superconductor transition.
This establishes an intimate relation between the magnetism and the
transport properties in the high-temperature copper oxide
superconductors.
Another important feature in lightly- and medium-doped
La$_{2-x}$Sr$_x$CuO$_4$ (0.04$\le x\le$0.12)
is that the charge density per unit length
estimated using a charge stripe model is almost constant throughout
the phase diagram, even when the modulation rotates away by 45$^\circ$
at the superconducting boundary. However, at a low value for
$x$ ($x$=0.024) the density appears to increase, approaching 1 hole/Cu as in
La$_{2-x}$Sr$_x$NiO$_4$.

It is unknown whether the stripes persist in very lightly-doped
La$_{2-x}$Sr$_x$CuO$_4$ ($x<$0.02) which shows coexistence of
a three-dimensional (3D) antiferromagnetic (AF) long-range ordered
phase and a spin-glass phase at low temperatures.
Several possibilities have been discussed theoretically. Some theories
predict that dilute holes in an antiferromagnet are unstable against
phase separation into a hole-rich and a hole-free phase.
\cite{emery,kivelson,hellberg,pryadko}
On the other hand, based on numerical calculations for a simple Hubbard
model, the incommensurability of the diagonal incommensurate peaks is
predicted to change proportionally with hole concentration.~\cite{kato}

Elastic neutron scattering measurements in
oxygenated La$_2$CuO$_4$ show that a magnetic Bragg peak, which
develops below the 3D AF transition temperature $T\rm_{N}\sim$90 K,
decreases in intensity below a re-entrant spin-glass order
temperature $T\rm_{m}\sim$25 K.~\cite{endoh,keimer} This is ascribed to
a depression of the 3D AF order because of the reentrant spin-glass
transition. Although it has been shown that two-dimensional (2D) AF
correlations develops below $T\rm_{m}$, no detailed study has been performed.

The present elastic neutron scattering study yields important new
information on the hole concentration dependence of
the magnetic properties in the coexistence region ($x<$0.02).
First we confirmed the reentrant behavior as described above.
It is found that in the reentrant phase diagonal incommensurate peaks
develop as in the pure spin-glass phase
in La$_{2-x}$Sr$_x$CuO$_4$ (0.02$\le x\le$0.055).
The integrated intensity of the diagonal incommensurate peaks that develop
below $T\rm_{m}$ corresponds closely to that of
the (1,0,0) magnetic Bragg peak lost below $T\rm_{m}$.
The most remarkable feature is that the parameters
(the incommensurability, the peak widths, and the transition temperature)
of the diagonal spin modulation at $x<$0.02 are locked at the values
in $x\sim$0.02 and only the volume fraction of the spin-glass phase
monotonically decreases with decreasing hole concentration ($c\rm_h$),
suggesting that the doped holes phase separate to form
some regions with $c\rm_h\sim$0.02 and the rest with $c\rm_h\sim$0.
Importantly, this behavior is similar to what was predicted for lightly-doped
holes in an antiferromagnet in the absence of long-range
Coulomb repulsion.~\cite{emery,kivelson,hellberg,pryadko}
\begin{figure}
\centerline{\epsfxsize=3.2in\epsfbox{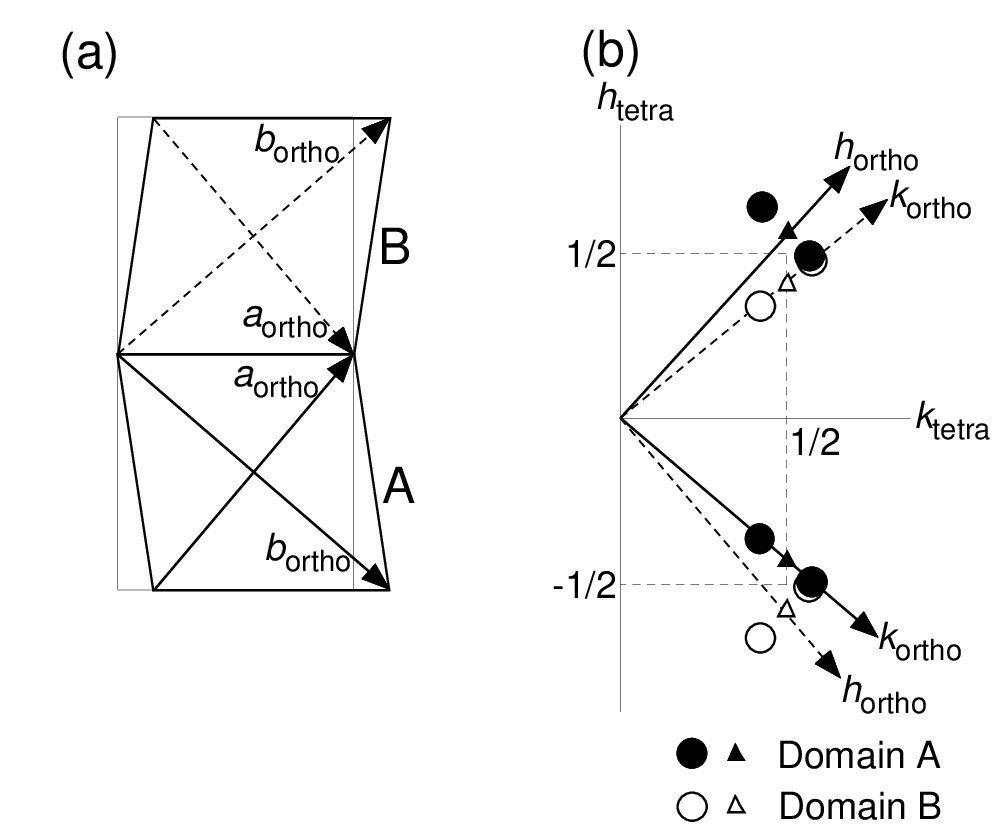}}
\caption{(a) A schematic drawing of the CuO$_2$ planes in the low
temperature orthorhombic structure. Two domains (A and B)
are shown. The thin lines represent the unit cell in the
high temperature tetragonal structure.
(b) Diagram of the reciprocal lattice in the $(HK0)$
scattering zone. Filled and open symbols are for domains A and B,
respectively. The circles and triangles correspond to the incommensurate
magnetic peaks and fundamental Bragg peaks, respectively.}
\label{fig1}
\end{figure}

\section{Experimental Details}
The single crystals of La$_{2-x}$Sr$_x$CuO$_4$ ($x$=0.01, 0.014, and
0.018) were grown by the traveling solvent floating zone (TSFZ) method.
The crystal was annealed in an Ar atmosphere at 900 $^\circ$C for 24 h.
From the 3D AF transition temperatures, the uncertainty in the effective hole
concentration of each crystal is estimated to be less than 10\%.
The twin structure of the crystals is shown in Fig. 1.
As shown in Fig. 2 of Ref. \onlinecite{wakimoto2}, four twins are
possible in the low temperature orthorhombic phase ($Bmab$).
A small number of twins greatly simplifies the analysis of
the incommensurate
magnetic peak structure. Fortunately, the $x$=0.010 and 0.014 crystals
have almost single domain structures. 
The $x$=0.018 crystal has two twins, which are estimated to be equally
distributed based on the ratio of the nuclear Bragg peak intensities
from both twins.

The neutron scattering experiments were carried out on the
cold neutron three-axis spectrometers LTAS and the thermal neutron
three-axis spectrometer TAS2 installed in the guide hall of JRR-3M at
the Japan Atomic Energy Research Institute (JAERI).
Typical horizontal collimator sequences were
guide-80-S-20$'$-80$'$ with a fixed incident neutron energy of
$E\rm_i$=5.05 meV at LTAS and guide-80$'$-S-40$'$-80$'$ with a fixed
incident neutron energy of $E\rm_i$=13.7 meV at TAS2.
Contamination from higher-order beams was effectively eliminated
using Be filters at LTAS and PG filters at TAS2.
The single crystal, which was oriented in the $(HK0)\rm_{ortho}$ or
$(H0L)\rm_{ortho}$ scattering plane, was mounted in a closed cycle
refrigerator.
In this paper, we use the low temperature
orthorhombic phase ($Bmab$) notation $(h,k,l)\rm_{ortho}$
to express Miller indices.
\begin{figure}
\centerline{\epsfxsize=3.2in\epsfbox{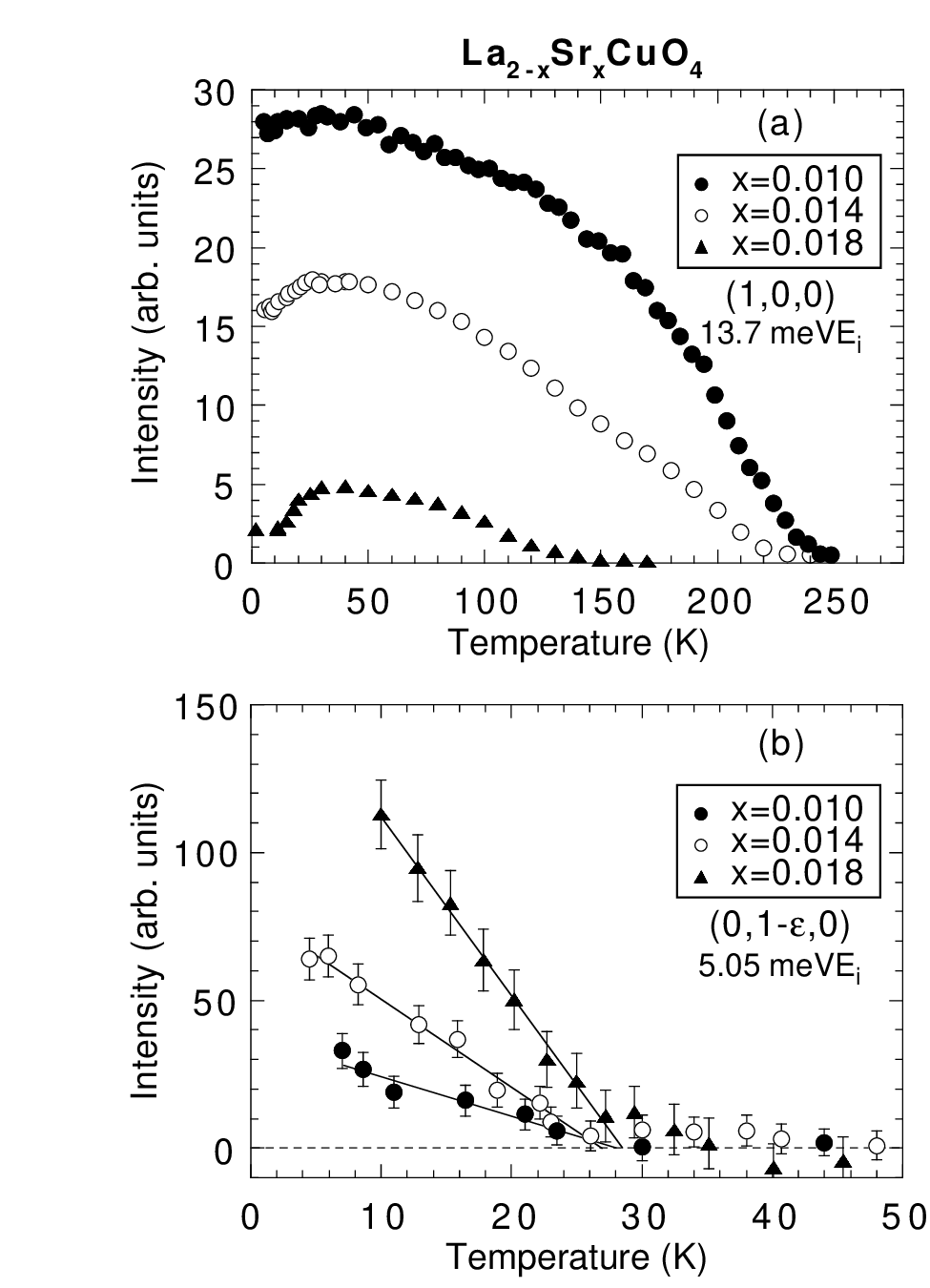}}
\caption{Temperature dependence of the (100) magnetic Bragg
intensity (a) and the magnetic intensity at the diagonal
incommensurate position (0,1-$\epsilon$,0) (b)
in La$_{2-x}$Sr$_x$CuO$_4$ ($x$=0.01, 0.014, and 0.018).
The solid lines are the results of fits to a linear function.
Background intensities measured at a high temperature have been
subtracted in (b).}
\label{fig2}
\end{figure}

\section{Results}
Figure 2(a) shows the temperature dependence of the (1,0,0) magnetic Bragg
intensities, originating from the 3D AF order, in La$_{2-x}$Sr$_x$CuO$_4$
($x$=0.01, 0.014, and 0.018). $T\rm_N$, as well as
the saturated intensity at $T\sim$30 K, corresponding to the square of
the ordered moment of the Cu$^{2+}$ ions, decrease with increasing
$c\rm_h$. 
This result is consistent with that in oxygenated
La$_2$CuO$_4$.~\cite{keimer,yamada0}
At a low temperature of $\sim$30 K, the magnetic Bragg intensities decrease
in the $x$=0.014 and 0.018 samples and at the same time
diagonal incommensurate peaks develop as in the pure spin-glass phase
in La$_{2-x}$Sr$_x$CuO$_4$ (0.02$\le x\le$0.055).
As shown in Fig. 2(b), diagonal incommensurate peaks develop
below $\sim$30 K in all of the samples.
The integrated intensity of the incommensurate peaks is calculated
using the same method as in Ref. \onlinecite{waki2}.
The integrated intensity that develops below $\sim$30 K corresponds
closely to that of the (1,0,0) magnetic Bragg peak lost below $\sim$30 K,
indicating that some regions of the 3D AF ordered phase turn into
the diagonal stripe phase.

The diagonal incommensurate peaks have been studied in more detail.
As shown in Fig. 2(a), the intense magnetic Bragg peak (1,0,0) still
remains below $T\rm_m$, which makes the measurement of the
incommensurate (1,$\pm\epsilon$,0) peaks difficult.
Therefore, the elastic measurements are performed around (0,1,0)
of domain A where no magnetic Bragg peak exists.
Although a small peak originating from a tail of the (1,0,0)
magnetic Bragg peak of the domain B exists near (0,1+$\epsilon$,0):
see the small open triangle in lower part of Fig. 1(b), the small
background peak can be easily removed by subtracting high temperature
data above $T\rm_{m}$.
Figure 3 shows elastic scans along $(0,K,0)$.
All of the samples clearly show the diagonal incommensurate structure.
The peak profiles differ between the $x$=0.010 and 0.014
samples and the $x$=0.018 sample. The two peaks are
symmetric in the $x$=0.010 and 0.014 samples because the domain structure
is almost single.
On the other hand, the two peaks are antisymmetric in the $x$=0.018
sample because of the nearly two-twin-structure.~\cite{peak}
The solid lines in Fig. 3 are the results of fits to a
convolution of the resolution function with 3D squared Lorentzians.
In the calculation one domain and equally distributed two twin
structure are assumed in the $x$=0.010 and $x$=0.018 sample, respectively.
A two twin structure with an unbalanced distribution is assumed in
the $x$=0.014 sample.
\begin{figure}
\centerline{\epsfxsize=3.0in\epsfbox{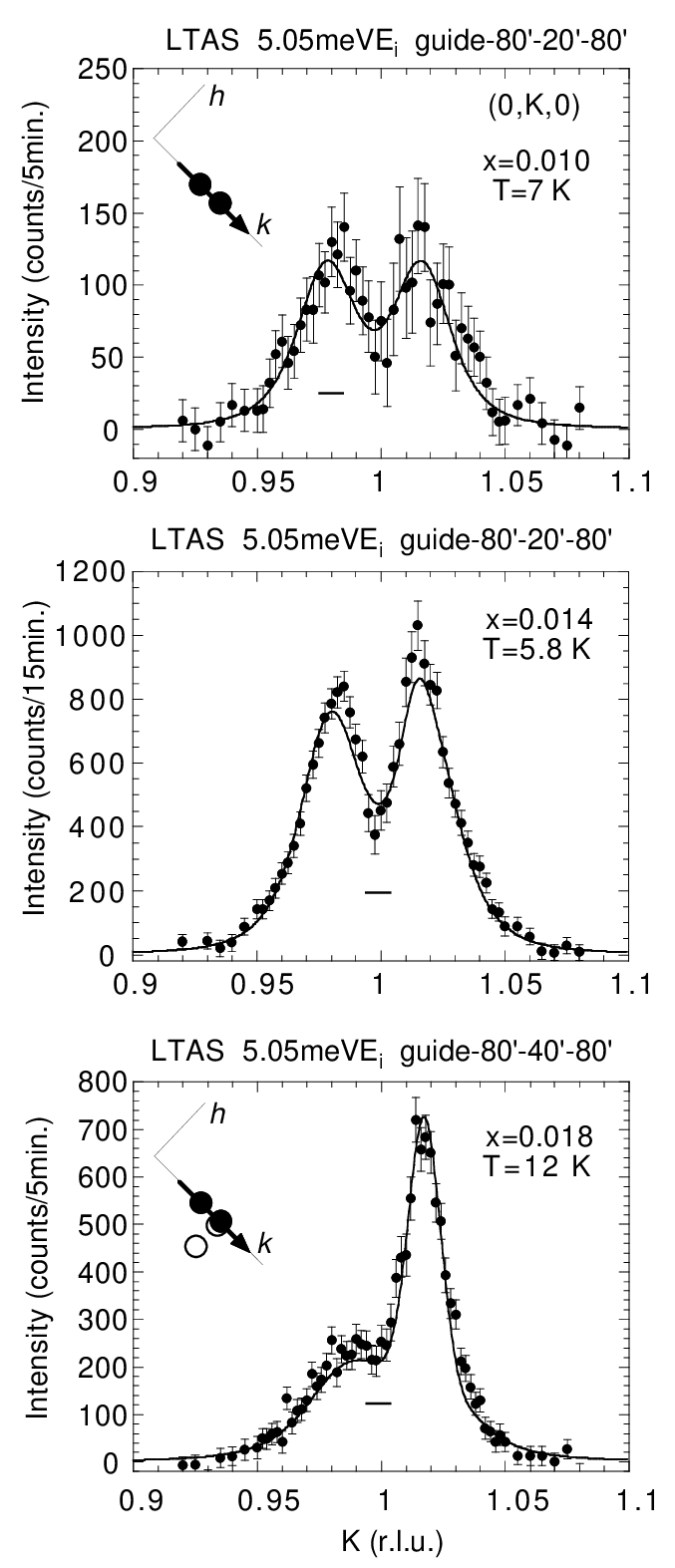}}
\caption{Elastic scans along $(0,K,0)$
in La$_{2-x}$Sr$_x$CuO$_4$ ($x$=0.01, 0.014, and 0.018).
The data measured above $T\rm_{m}$ are subtracted as background
intensities.
The solid lines are the results of fits to a convolution of
the resolution function with 3D squared Lorentzians
with the parameters shown in Table 1.
The small horizontal bar in each figure represents the instrumental
resolution full width.
The insets show the magnetic peak positions around (0,1,0) in the $(HK0)$
scattering plane. The thick arrows show scan trajectories.}
\label{fig3}
\end{figure}
\begin{figure}
\centerline{\epsfxsize=3.2in\epsfbox{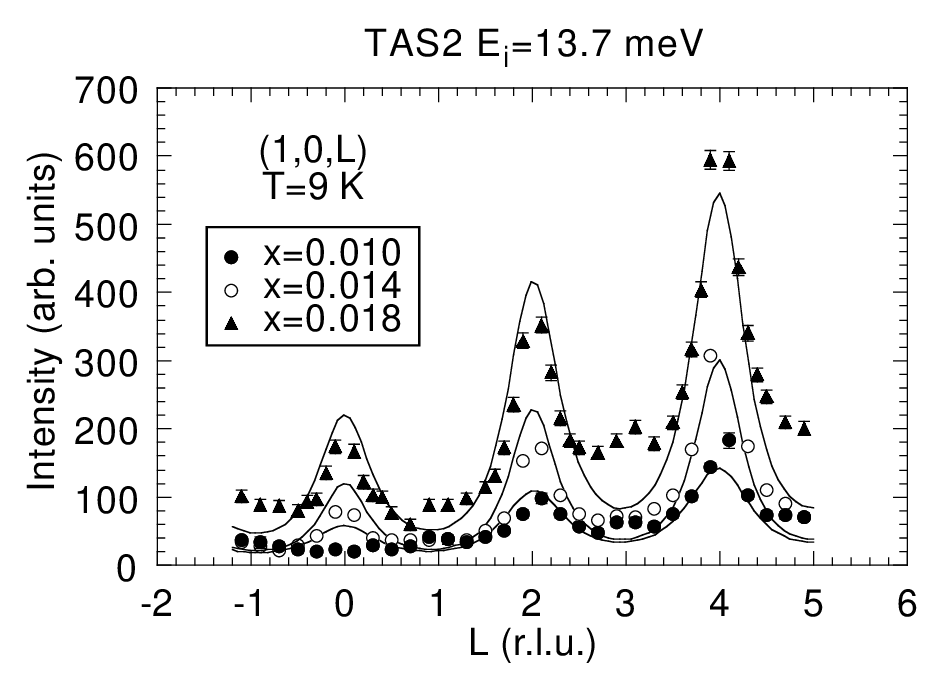}}
\caption{Elastic scans along $(1,0,L)$ at $T$=9 K
in La$_{2-x}$Sr$_x$CuO$_4$ ($x$=0.01, 0.014, and 0.018).
Background intensities measured at a high temperature have been
subtracted.
The solid lines are the results of fits to a convolution of
the resolution function with 3D squared Lorentzians
with the parameters shown in Table 1.}
\label{fig4}
\end{figure}

Figure 4 shows elastic scans along $(1,0,L)$ perpendicular to the
CuO$_2$ planes in the three samples.
The solid lines in Fig. 4 are the calculated profiles using
3D squared Lorentzian profiles convoluted with the instrumental
resolution function. 
In order to reproduce the $L$-dependence of the $(1,0,even)$ intensity,
the cluster spin-glass model \cite{matsuda}
has been used in the calculation.
The calculation describes the observed profiles in the $(HK0)$ and
$(H0L)$ zone reasonably well.
The fitted parameters are listed in Table 1. The peak profile is anisotropic
as in the $x$=0.024 sample, that is, in the pure spin-glass phase.
One characteristic feature is that the peak is very sharp and close
to resolution-limited along the $a$ axis, which is the direction along
which the stripes run.
The large inverse peak widths of the incommensurate peaks
presumably reflect the fact that the background of the diagonal
stripe phase is the long-range 3D AF phase, that is, the long-range AF
correlations enhance the correlation length of the diagonal stripe phase.
Another characteristic feature is that the inverse peak widths
($\xi\rm_a >$500 \AA, $\xi\rm_b\sim$200 \AA, $\xi\rm_c\sim$10
\AA)
and the incommensurability ($\sim$0.019 r.l.u.)
do not depend on hole concentration below $x<$0.02.~\cite{error}

A summary of the elastic neutron scattering results is shown in Figs. 5
and 6.
Fig. 5(a) shows the hole concentration dependence of
$T\rm_{N}$ and $T\rm_{m}$,
where $T\rm_{m}$ is calculated assuming that the intensity increases 
linearly below the temperature.
$T\rm_{N}$ gradually decreases with increasing $c\rm_h$.
On the other hand, $T\rm_{m}$ is constant at
$\sim$28 K for $x<$0.02 and decreases with increasing $c\rm_h$
at $x>$0.02. As shown in Fig. 6, the incommensurability $\epsilon$ is
locked around 0.02 below $x$=0.02 and then follows the relation
$\epsilon$=$x$ in a narrow region around $x\sim$0.02 and finally
follows the relation $\epsilon >x$ above $x$=0.03.~\cite{eps}
The incommensurability corresponds to the inverse modulation period
of the spin density wave.
In the diagonal charge stripe model, the relation $\epsilon$=$x$
corresponds to a constant charge per unit length 1 hole/Cu.
On the other hand, the relation $\delta =x$, where $\delta$ is
in the high temperature tetragonal notation and
$\epsilon=\sqrt{2}\times\delta$, corresponds to 0.7 hole/Cu,
indicating that some of the doped holes are located between
the charge stripes.
The ordered moment of the Cu$^{2+}$ ions in the 3D AF phase at 30 K quickly
decreases with increasing $c\rm_h$ as shown in Fig. 5(b).
The Cu$^{2+}$ moment in the diagonal spin-glass phase
below $\sim$30 K is also plotted in the same figure. If
the moment is proportional to $x$, it is extrapolated to be
$\sim$0.1$\mu\rm_B$ at $x$=0.02.
Assuming that the Cu$^{2+}$ moment is 0.1$\mu\rm_B$ in the spin-glass
phase of the coexistence region, the volume fraction of the spin-glass
phase can be estimated as in the inset of Fig. 5(b).
\begin{figure}
\centerline{\epsfxsize=3.2in\epsfbox{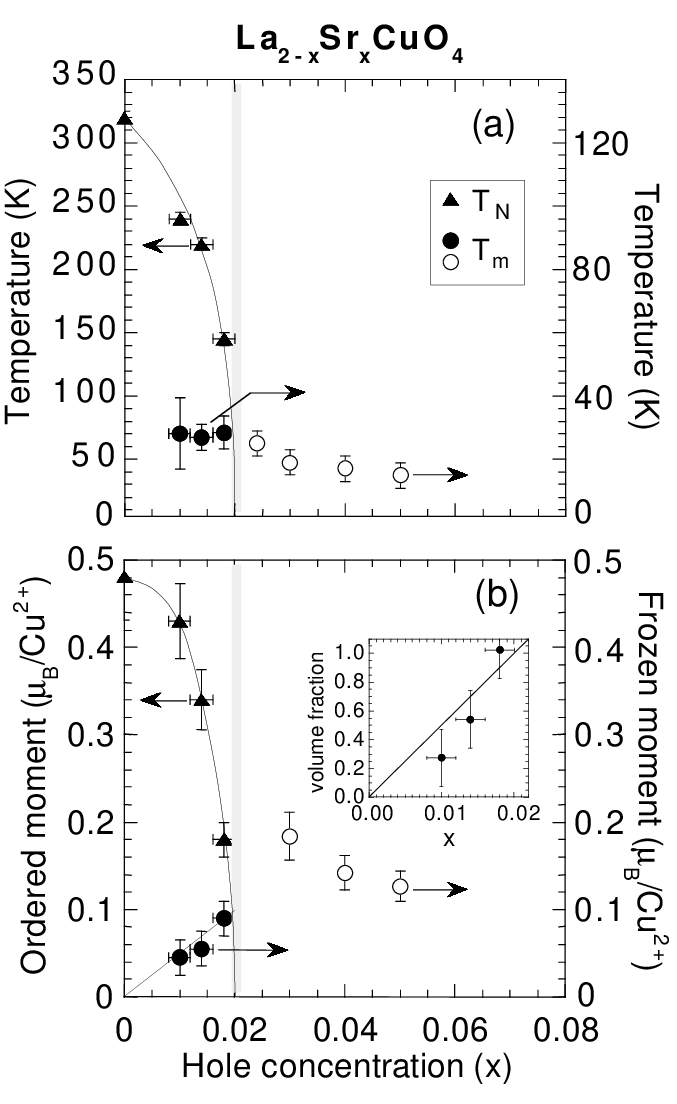}}
\caption{Hole concentration ($x$) dependence of the magnetic
transition temperatures (a) and ordered moment at $T$=30 K and
frozen moment at $T$=4 K(b).
$T\rm_{N}$ and $T\rm_{m}$ in (a) represent the 3D AF transition temperature
and the spin-glass transition temperature, respectively.
The inset in (b) shows the estimated
volume fraction of the spin-glass phase.
Filled symbols in (a) and (b) are determined from the present study.
Open symbols in (a) and (b) are from Refs. 4 and 19, respectively.
The solid lines are guides-to-the-eyes.}
\label{fig5}
\end{figure}
\begin{figure}
\centerline{\epsfxsize=3.2in\epsfbox{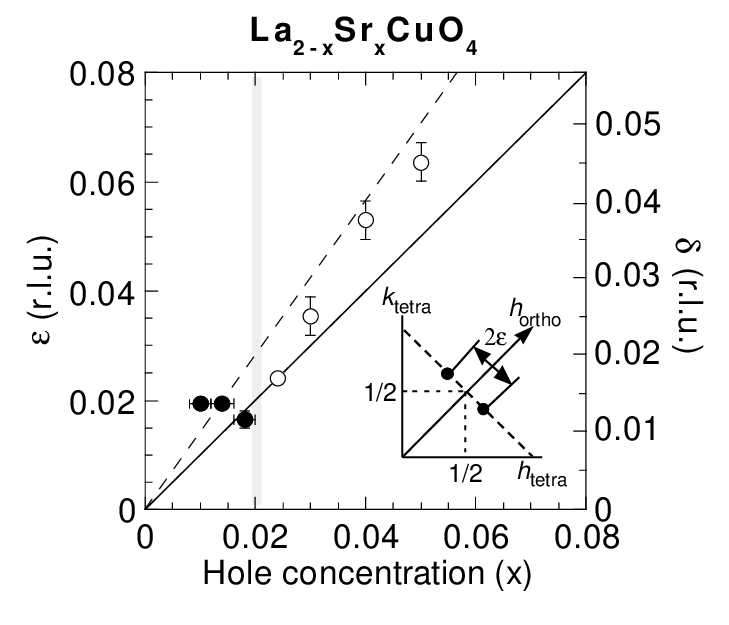}}
\caption{Hole concentration ($x$) dependence of the splitting of the
incommensurate peaks.
The inset shows the configuration of the incommensurate peaks
in the diagonal stripe phase.
Filled symbols are determined from the present study.
Open symbols are from Refs. 2 and 3.
$\epsilon=\sqrt{2}\times\delta$ where $\delta$ is defined in
tetragonal units.
The solid and broken lines correspond to $\epsilon =x$ and
$\delta = x$, respectively.}
\label{fig6}
\end{figure}

\section{Discussion and Conclusions}
As shown above, some part of the 3D AF ordered phase turns into the
spin-glass phase, in which the stripe structure is almost identical to
that at $x\sim$0.02.
These results indicate that the doped holes
phase separate microscopically to form finite size regions of
the $x$=0.02 stripe phase. Therefore, the hole concentration
in the rest of the regions must be much less than $x$.
If the volume fraction of the spin-glass phase is proportional to $x$,
the coexistence phase is a mixture of spin-glass phase
with $c\rm_h\sim$0.02 and 3D AF phase with $c\rm_h$ close to 0.
These results suggest that the transition between the 3D AF and spin-glass
phases around $x$=0.02 is first order and there exists a miscibility
gap to form the diagonal stripe. This behavior is predicted for a
doped antiferromagnet in the absence of a long-range Coulomb interaction.
\cite{emery,kivelson,hellberg,pryadko}
In this study, we determined that $c\rm_h\sim$0.02 at the hole-rich region,
which was not specified in the theoretical studies.
The behavior is much different from that expected from the theory
by Kato $et$ $al.$,~\cite{kato}
which has predicted that the incommensurability should follow the relation
$\epsilon$=$x$ down to very low hole concentrations.
Since it is believed that the orthorhombic distortion ($Bmab$) stabilizes
the diagonal stripe structure, the crystal structure should favor the stripe
structure in the lower hole concentration region. Furthermore, any chemical
or structural disorder originating from the Sr doping will decrease
with decreasing $x$.
Therefore, the electronic phase separation of the doped holes
must be an intrinsic phenomenon.
It is noted that parallel stripes, observed in superconducting
La$_{2-x}$Sr$_x$CuO$_4$ ($x>$0.06), are the most stable at $x\sim$0.12
and the incommensurability saturates at
$x>$0.12.~\cite{tranquada2,yamada} However, no electronic phase separation
seems to occur in the $x>$0.12 hole concentration region but rather the
charge density between charge stripes becomes larger in order to
stabilize the stripe periodicity.

We now compare the results of our magnetic study described above with
those of previous transport studies.~\cite{preyer,ando}
It is reported that even in lightly-doped La$_{2-x}$Sr$_x$CuO$_4$
(0.01$\le x\le$0.03) the in-plane conductivity shows metallic behavior
above $T\rm_{L}\sim$50-150 K. At temperatures below $T\rm_{L}$ the doped holes
apparently start to localize.
Magnetically, our results show that a reentrant
spin-glass phase with the diagonal incommensurate structure appears
at $\sim$30 K.
Accordingly, as shown above, hole localization does not seem to occur
homogeneously but rather the doped holes form finite size regions with
$c\rm_h\sim$0 or 0.02 for $T<$30 K.

We now compare the results of our neutron scattering study with
those of muon spin relaxation ($\mu$SR) and nuclear quadrupole
resonance (NQR) studies.~\cite{borsa,chou}
It has been found that a static internal field, determined from both
$^{139}$La NQR
and $\mu$SR, develops below $T\rm_{N}$ with a further increase at
$T\sim$30 K, and the latter temperature does not depend on the hole
concentration.
The two characteristic temperatures are consistent with those
determined from our neutron scattering studies.
Furthermore, the internal field at $T$=0 K appears to be independent of
the hole concentration, suggesting that the ordered moment is also
independent of hole concentration. This is consistent with our result
that the hole concentration of the 3D AF phase becomes almost zero
at $T$=0 K as shown above.
On the other hand, the NQR studies show that the spin-glass transition
temperature $T\rm_f$, determined from the occurrence of a sharp peak
in the nuclear spin-lattice relaxation rate as a function of temperature,
follows the relation $T\rm_f$$\propto$$x$ below $x<$0.02.
This result is different from that in the neutron scattering studies.
One possibility is that $T\rm_f$ depends directly on the volume fraction
of the spin-glass phase since our neutron scattering studies show
that the volume fraction of the spin-glass phase decrease almost linearly
with decreasing $c\rm_h$.~\cite{note}
\begin{table}
\caption{Hole concentration dependence of the inverse peak
widths in La$_{2-x}$Sr$_x$CuO$_4$.}
\label{table1}
\begin{tabular}{cccc}
$x$ & $\xi'_a$ (\AA) & $\xi'_b$ (\AA) & $\xi'_c$ (\AA)\\ \tableline
0.010 & $>$500 & 170(10) & 8(2)\\
0.014 & $>$500 & 200(10) & 9(2)\\
0.018 & $>$500 & 170(20) & 11(1)\\
0.024\tablenote{Ref. 3.} & 95(4) & 40(1) & 3.15(8)
\end{tabular}
\end{table}

In summary, our neutron scattering experiments demonstrate
the electronic phase separation of the doped holes in
lightly-doped La$_{2-x}$Sr$_x$CuO$_4$, which is predicted for a doped
antiferromagnet. Some clusters
with $c\rm_h\sim$0.02 exhibit diagonal stripe correlations
while the rest of the crystal with $c\rm_h\sim$0 shows 3D AF order.

\section*{Acknowledgments}
We would like to thank K. Ishida and S. Wakimoto for
stimulating discussions. This study was supported in part by the
U.S.-Japan Cooperative Program on Neutron Scattering, by a Grant-in-Aid
for Scientific Research from the Japanese Ministry of Education, Science,
Sports and Culture, by a Grant for the Promotion of Science from the
Science and Technology Agency, and by CREST.
Work at Brookhaven National Laboratory was carried out under Contract
No. DE-AC02-98CH10886, Division of Material Science, U.S. Department of
Energy. Work at the University of Toronto is part of the Canadian
Institute for Advanced Research and is supported by the Natural Science
and Engineering Research Council of Canada.


\begin{references}
\bibitem{wakimoto}S. Wakimoto, R. J. Birgeneau, Y. Endoh, P. M. Gehring,
K. Hirota, M. A. Kastner, S. H. Lee, Y. S. Lee, G. Shirane, S. Ueki,
and K. Yamada, Phys. Rev. B {\bf60}, R769 (1999).
\bibitem{wakimoto2}S. Wakimoto, R. J. Birgeneau, M. A. Kastner,
Y. S. Lee, R. Erwin, P. M. Gehring, S. H. Lee, M. Fujita, K. Yamada,
Y. Endoh, K. Hirota, and G. Shirane, Phys. Rev. B {\bf61}, 3699 (2000).
\bibitem{matsudanew} M. Matsuda, M. Fujita, K. Yamada, R. J. Birgeneau,
M. A. Kastner, H. Hiraka, Y. Endoh, S. Wakimoto, and G. Shirane,
Phys. Rev. B {\bf62}, 9148 (2000).
\bibitem{fujita}M. Fujita, K. Yamada, H. Hiraka, P. M. Gehring,
S. H. Lee, S. Wakimoto, and G. Shirane, Phys. Rev. B {\bf65}, 64505
(2002).
\bibitem{machida0}K. Machida, Physica C {\bf158}, 192 (1989).
\bibitem{kato}M. Kato, K. Machida, H. Nakanishi, and M. Fujita,
J. Phys. Soc. Jpn. {\bf59}, 1047 (1990).
\bibitem{rice}D. Poilblanc and T. M. Rice, Phys. Rev. B {\bf39}, 9749
(1989).
\bibitem{schulz}H. Schulz, J. Phys. (Paris) {\bf50}, 2833 (1989).
\bibitem{zaanen}J. Zaanen and O. Gunnarsson, Phys. Rev. B {\bf40}, 7391
(1990).
\bibitem{tranquada3}J. M. Tranquada, D. J. Buttrey, V. Sachan,
Phys. Rev. B {\bf54}, 12318 (1996).
\bibitem{yoshizawa}H. Yoshizawa, T. Kakeshita, R. Kajimoto, T. Tanabe,
T. Katsufuji, and Y. Tokura, Phys. Rev. B {\bf61}, R854 (2000).
\bibitem{emery}V. J. Emery, S. A. Kivelson, and H. Q. Lin,
Phys. Rev. Lett. {\bf64}, 475 (1990).
\bibitem{kivelson}S. A. Kivelson and V. J. Emery, in $Proceedings$ $of$
$the$ $Los$ $Alamos$ $Symposium-1993:$ $Strongly$ $Correlated$
$Electronic$ $Materials$, edited by K. S. Bedell $et$ $al$.
(Addison-Weley, NY, 1994), p. 619.
\bibitem{hellberg}C. S Hellberg and E. Manousakis,
Phys. Rev. Lett. {\bf78}, 4609 (1997).
\bibitem{pryadko}L. P. Pryadko, S. Kivelson, and D. W. Hone,
Phys. Rev. Lett. {\bf80}, 5651 (1998).
\bibitem{endoh}Y. Endoh, K. Yamada, R. J. Birgeneau, D. R. Gabbe,
H. P. Jenssen, M. A. Kastner, C. J. Peters, P. J. Picone,
T. R. Thurston, J. M. Tranquada, G. Shirane, Y. Hidaka, M. Oda,
Y. Enomoto, M. Suzuki, and T. Murakami, Phys. Rev. B {\bf37}, 7443
(1988).
\bibitem{keimer}B. Keimer, N. Belk, R. J. Birgeneau, A. Cassanho, C. Y. Chen,
M. Greven, M. A. Kastner, A. Aharony, Y. Endoh, R. W. Erwin, and G. Shirane,
Phys. Rev. B {\bf46}, 14034 (1992).
\bibitem{yamada0}K. Yamada, E. Kudo, Y. Endoh, Y. Hidaka, M. Oda,
M. Suzuki, and T. Murakami, Solid state Commun. {\bf64}, 753 (1987).
\bibitem{waki2} S. Wakimoto, R. J. Birgeneau, Y. S. Lee, and G. Shirane,
Phys. Rev. B {\bf63}, 172501 (2001).
\bibitem{peak} Detailed description of the peak profile in the twin structure
is shown in Ref. 3.
\bibitem{matsuda}M. Matsuda, R. J. Birgeneau, P. B\"{o}ni,
Y. Endoh, M. Greven, M. A. Kastner, S.-H. Lee, Y. S. Lee,
G. Shirane, S. Wakimoto, K. Yamada, Phys. Rev. B {\bf61}, 4326 (2000).
\bibitem{error}The error for the incommensurability is large
in the $x$=0.018 sample because the peak splitting along $K$ is
difficult to resolve due to the twin structure.
\bibitem{eps}The error for the incommensurability becomes smaller at
lower hole concentrations because the peaks become sharper as
shown in Table 1.
\bibitem{borsa}F. Borsa, P. Carretta, J. H. Cho, F. C. Chou, Q. Hu,
D. C. Johnston, A. Lascialfari, D. R. Torgeson, R. J. Gooding,
N. M. Salem, and K. J. E. Vos, Phys. Rev. B {\bf52}, 7334
(1995).
\bibitem{tranquada2}J. M. Tranquada, J. D. Axe, N. Ichikawa,
Y. Nakamura, S. Uchida, and B. Nachumi, Phys. Rev. B {\bf54}, 7489 (1996).
\bibitem{yamada}K. Yamada, C. H. Lee, K. Kurahashi, J. Wada, S. Wakimoto,
S. Ueki, H. Kimura, Y. Endoh, S. Hosoya, G. Shirane, R. J. Birgeneau,
M. Greven, M. A. Kastner, and Y. J. Kim,
Phys. Rev. B {\bf57}, 6165 (1998).
\bibitem{preyer}N. W. Preyer, M. A. Kastner, C. Y. Chen, R. J. Birgeneau,
and Y. Hidaka, Phys. Rev. B {\bf44}, 407 (1991).
\bibitem{ando}Y. Ando, A. N. Lavrov, S. Komiya, and X. F. Sun,
Phys. Rev. Lett. {\bf87}, 17001 (2001).
\bibitem{chou} F. C. Chou, F. Borsa, J. H. Cho, D. C. Johnston,
A. Lascialfari, D. R. Torgeson, and J. Ziolo,
Phys. Rev. Lett. {\bf71}, 2323 (1993).
\bibitem{note} The characteristic lengths in Table 1 do not necessarily
represent the cluster size of the spin-glass phase. As suggested in Ref. 3,
the incommensurate magnetic peaks could be broad because of disorder
in both the periodicity and the direction of the stripes.
Therefore, those lengths represent the minimum sizes for the clusters.
\end{references}
\end{document}